\documentclass[aps,onecolumn,showpacs,nofootinbib,showkeys]{revtex4-2}
\usepackage[margin=2.15cm]{geometry}

\RequirePackage[T1]{fontenc}

\usepackage{epsfig,graphicx,amsmath,amssymb,bm}
\RequirePackage{mathptmx}      
\usepackage{mathrsfs}
\usepackage[english]{babel}
\usepackage{amssymb}
\RequirePackage{color}
\usepackage{changes}
\usepackage{slashed}
\usepackage{multirow}
\usepackage{scalerel}
\usepackage{tikz-feynman}
\usepackage{textcomp}
\usepackage{subcaption}
\usepackage[english]{babel}
\usepackage{float}
\RequirePackage{hyperref}
\hypersetup{
    linktocpage,
    colorlinks,
    citecolor=blue,
    filecolor=black,
    linkcolor=blue,
    urlcolor=blue,
}
\usepackage{nccmath}

\begin{document}

\title{Decays $\tau \to 3K \nu_\tau$ in $U(3)\times U(3)$ quark NJL model}


\author{M.K. Volkov$^{1}$}\email{volkov@theor.jinr.ru}
\author{A.A. Pivovarov$^{1}$}\email{tex$\_$k@mail.ru}
\author{K. Nurlan$^{1,2,3}$}\email{nurlan@theor.jinr.ru}

\affiliation{$^1$ Bogolubov laboratory of theoretical physics, Joint institute for nuclear research, Dubna, 141980 Russia \\
                $^2$ Institute of Nuclear Physics, Almaty,, 050032, Kazakhstan \\
                $^3$ Eurasian National University, Astana, 010008, Kazakhstan}   


\begin{abstract}
The widths of the decays $\tau \to K^- K^+ K^- \nu_\tau$ and $\tau \to K^- K^0 \bar{K}^0 \nu_\tau$ are calculated in the $U(3)\times U(3)$ chiral quark NJL model. Four channels are considered: contact, axial vector, vector and pseudoscalar channels. It is shown that the dominant contribution is given by the axial vector channel with an intermediate $\phi$ meson. The obtained results are in satisfactory agreement with the experimental data.  


\end{abstract}


\maketitle


\section{\label{Intro}Introduction}
The widths of the decay $\tau \to K^- K^+ K^- \nu_\tau$ measured by various collaborations are differ significantly from each other. The BaBar collaboration obtained for the branching fraction the value $(1.58 \pm 0.25) \times 10^{-5}$ \cite{BaBar:2007chl}. At the same time, the Belle collaboration obtained twice the value $(3.29 \pm 0.37) \times 10^{-5}$ \cite{Belle:2010fal}. Such a discrepancy in the experimental data lends special interest in theoretical research of this decay. Particularly, the main attention should be paid to the internal structure of this process. The corresponding research requires the application of a phenomenological model relevant to the description of hadron interactions at low energies.

One of the most effective models of this type allowing to describe the hadron decays of the $\tau$ lepton is the Nambu--Jona-Lasinio (NJL) model \cite{Nambu:1961tp, Eguchi:1976iz, Ebert:1982pk, Volkov:1984kq, Volkov:1986zb, Ebert:1985kz, Vogl:1991qt, Klevansky:1992qe, Volkov:1993jw, Hatsuda:1994pi, Ebert:1994mf, Volkov:2005kw}. According to a series of our previous studies, the application of this model entails quite satisfactory results in the description of such processes~\cite{Volkov:2017arr, Volkov:2022jfr}.

In the present work, the processes $\tau \to K^- K^+ K^- \nu_\tau$ and $\tau \to K^- K^0 \bar{K}^0 \nu_\tau$ are described theoretically in the framework of the NJL model. In calculation of the branching fractions of these decays none of the additional arbitrary parameters are used. This research provide a deeper understanding of the internal structure and mechanisms of such processes.
       
\section{The Lagrangian of the NJL model}
For calculation of the widths of the considered processes we need the following vertices of the quark-meson Lagrangian of the NJL model~\cite{Volkov:2005kw,Volkov:2022jfr}:
\begin{eqnarray}
	\Delta L_{int} & = & \bar{q}\left\{\sum_{i=0,\pm}\left[ig_{K}\gamma^{5}\lambda^{K}_{i}K^{i} + \frac{g_{\rho}}{2}\gamma^{\mu}\lambda^{\rho}_{i}\rho^{i}_{\mu} + \frac{g_{K^{*}}}{2}\gamma^{\mu}\lambda^{K}_{i}K^{*i}_{\mu} + \frac{g_{K_1}}{2}\gamma^{\mu}\gamma^{5}\lambda^{K}_{i}K^{i}_{1A\mu} \right]\right. \nonumber\\
	&&\left. + ig_{K}\gamma^{5}\lambda_{0}^{\bar{K}}\bar{K}^{0} + \frac{g_{K^{*}}}{2}\gamma^{\mu}\lambda^{\bar{K}}_{0}\bar{K}^{*0}_{\mu} + \frac{g_{\omega}}{2}\gamma^{\mu}\lambda^{\omega}\omega_{\mu} + \frac{g_{\phi}}{2}\gamma^{\mu}\lambda^{\phi}\phi_{\mu}\right\}q,
\end{eqnarray}
where $q$ and $\bar{q}$ are the u-, d- and s-quark fields with the constituent masses $m_{u} = m_{d} = 270$~MeV, $m_{s} = 420$~MeV and $\lambda$ are the linear combinations of the Gell-Mann matrices.

The strange axial vector meson $K_{1A}$ is presented as a sum of two physical states~\cite{Volkov:1984gqw, Suzuki:1993yc, Volkov:2019awd}:
\begin{eqnarray}
    K_{1A} = K_1(1270)\sin{\alpha} + K_1(1400)\cos{\alpha},
\end{eqnarray}
where $\alpha = 57^\circ$.

The coupling constants:
\begin{displaymath}
	g_{\rho} = g_{\omega} = \sqrt{\frac{3}{2 I_{20}}}, \quad g_{\phi} = \sqrt{\frac{3}{2 I_{02}}}, \quad g_{K} = \sqrt{\frac{Z_{K}}{4 I_{11}}}, \quad g_{K^{*}} = g_{K_1} = \sqrt{\frac{3}{2 I_{11}}},
\end{displaymath}
where
\begin{eqnarray}
	&Z_{K} = \left(1 - \frac{3}{2}\frac{(m_{u} + m_{s})^{2}}{M^{2}_{K_{1A}}}\right)^{-1},& \nonumber\\
	&M^{2}_{K_{1A}} = \left(\frac{\sin^{2}{\alpha}}{M^{2}_{K_{1}(1270)}} - \frac{\cos^{2}{\alpha}}{M^{2}_{K_{1}(1400)}}\right)^{-1},&
\end{eqnarray}
$Z_{K}$ is the factor appearing when the $K - K_{1}$ transition is taken into account, $M_{a_{1}} = 1230$~MeV, $M_{K_{1}(1270)} = 1253$~MeV, $M_{K_{1}(1400)} = 1403$~MeV~\cite{ParticleDataGroup:2022pth} are the masses of the axial vector mesons $a_{1}$ and $K_{1}$.

The integrals used in the definitions of the coupling constants and appearing in the quark loops in the renormalization of the Lagrangian take the form
\begin{equation}
\label{integral}
	I_{nm} = -i\frac{N_{c}}{(2\pi)^{4}}\int\frac{\theta(\Lambda^{2} + k^2)}{(m_{u}^{2} - k^2)^{n}(m_{s}^{2} - k^2)^{m}}
	\mathrm{d}^{4}k,
\end{equation}
where $\Lambda = 1265$ MeV is the cut-off parameter~\cite{Volkov:2022jfr}.

\section{Decay $\tau \to K^- K^+ K^- \nu_\tau$}
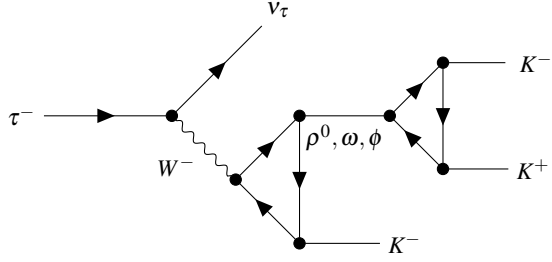
\begin{figure*}[t]
 \centering
  \begin{subfigure}{0.5\textwidth}
   \centering
    \begin{tikzpicture}
     \begin{feynman}
      \vertex (a) {\(\tau^-\)};
      \vertex [dot, right=2cm of a] (b){};
      \vertex [above right=2cm of b] (c) {\(\nu_{\tau}\)};
      \vertex [dot, below right=1.2cm of b] (d) {};
      \vertex [dot, above right=1.2cm of d] (e) {};
      \vertex [dot, below right=1.2cm of d] (h) {};
      \vertex [dot, right=1.2cm of e] (f) {};
      \vertex [dot, above right=1.0cm of f] (n) {};  
      \vertex [dot, below right=1.0cm of f] (m) {};   
      \vertex [right=1.2cm of n] (l) {\(\ K^- \)}; 
      \vertex [right=1.2cm of m] (s) {\(K^+\)};  
      \vertex [right=1.4cm of h] (k) {\(K^- \)}; 
      \diagram* {
         (a) -- [fermion] (b),
         (b) -- [fermion] (c),
         (b) -- [boson, edge label'=\(W^-\)] (d),
         (d) -- [fermion] (e),  
         (e) -- [fermion] (h),
         (d) -- [anti fermion] (h),
         (e) -- [edge label'=\({ \rho^0, \omega, \phi} \)] (f),
         (f) -- [fermion] (n),
         (n) -- [fermion] (m),
         (f) -- [anti fermion] (m), 
         (h) -- [] (k),
         (n) -- [] (l),
	 (m) -- [] (s),
      };
     \end{feynman}
    \end{tikzpicture}
  \end{subfigure}%
 \caption{The contact diagram of the decay $\tau \to K^- K^+ K^- \nu_\tau$.}
 \label{diagram1}
\end{figure*}

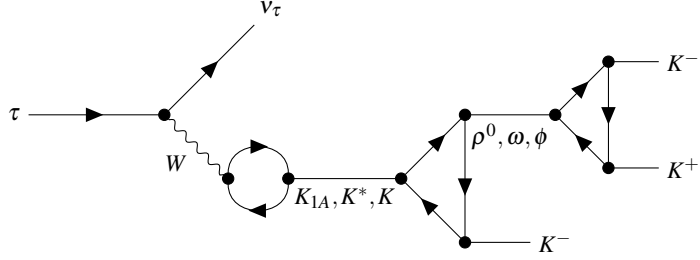
\begin{figure*}[t]
 \centering
 \centering
 \begin{subfigure}{0.5\textwidth}
  \centering
   \begin{tikzpicture}
    \begin{feynman}
      \vertex (a) {\(\tau\)};
      \vertex [dot, right=2cm of a] (b){};
      \vertex [above right=2cm of b] (c) {\(\nu_{\tau}\)};
      \vertex [dot, below right=1.2cm of b] (d) {};
      \vertex [dot, right=0.8cm of d] (l) {};
      \vertex [dot, right=1.5cm of l] (g) {};
      \vertex [dot, above right=1.2cm of g] (e) {};
      \vertex [dot, below right=1.2cm of g] (h) {};      
      \vertex [dot, right=1.2cm of e] (f) {};
      \vertex [dot, above right=1.0cm of f] (n) {};
      \vertex [dot, below right=1.0cm of f] (m) {};
      \vertex [right=1.0cm of n] (s) {\( K^- \)};
      \vertex [right=1.0cm of m] (r) {\( K^+ \)};
      \vertex [right=1.2cm of h] (k) {\( K^- \)}; 
      \diagram* {
         (a) -- [fermion] (b),
         (b) -- [fermion] (c),
         (b) -- [boson, edge label'=\(W\)] (d),
         (d) -- [fermion, inner sep=1pt, half left] (l),
         (l) -- [fermion, inner sep=1pt, half left] (d),
         (l) -- [edge label'=\({ K_{1A}, K^{*}, K } \)] (g),
         (g) -- [anti fermion] (h),  
         (h) -- [anti fermion] (e),
         (e) -- [anti fermion] (g),      
         (e) -- [edge label'=\( { \rho^0, \omega, \phi } \)] (f),
         (f) -- [fermion] (n),
         (n) -- [fermion] (m),
         (m) -- [fermion] (f),
         (h) -- [] (k),
         (n) -- [] (s),
         (m) -- [] (r),
      };
     \end{feynman}
    \end{tikzpicture}
  \end{subfigure}%
 \caption{The diagram with the intermediate mesons describing the decay $\tau \to K^- K^+ K^- \nu_\tau$.}
 \label{diagram2}
\end{figure*}

The diagrams describing the decay $\tau \to K^- K^+ K^- \nu_\tau$ are presented in Figs.~\ref{diagram1} and \ref{diagram2}.

The amplitude of this process takes the form
\begin{equation}
\label{amplitude}
    \mathcal{M} = G_{F} V_{us} L_{\mu} \left\{\mathcal{M}_{A} + \mathcal{M}_{V} + \mathcal{M}_{P}\right\}^{\mu},
\end{equation}
where $L_{\mu}$ is the lepton current, $\mathcal{M}_{A}$, $\mathcal{M}_{V}$ and $\mathcal{M}_{P}$ are the axial vector, vector and pseudoscalar channels:
\begin{eqnarray}
    \mathcal{M}_{A}^{\mu} & = & \frac{i}{8} (m_{s} + m_{u}) \frac{Z_{K}}{g_{K}} \left[BW_{K_1(1270)} \sin{\alpha} + BW_{K_1(1400)} \cos{\alpha}\right]\left[g^{\mu\nu}h_{A} - q^{\mu}q^{\nu}\right] \nonumber\\
    && \times\left\{g_{\rho}^2 BW_{\rho^0} + g_{\omega}^2 BW_{\omega} + 2g_{\phi}^2BW_{\phi}\right\}\left(p_{K^+} - p_{K^-}^{(1)}\right)_{\nu} + \left(p_{K^-}^{(1)} \leftrightarrow p_{K^-}^{(2)}\right), \nonumber\\
    \mathcal{M}_{V}^{\mu} & = & 2 m_{u} g_{K} I_{c} h_{V} BW_{K^{*-}} \left\{g_{\rho}^{2} BW_{\rho^0} + g_{\omega}^{2} BW_{\omega} - 2 g_{\phi}^{2} BW_{\phi} \right\} \nonumber\\
    && \times e^{\mu \nu \lambda \delta} p_{K^+\nu} p_{K^{-}\lambda}^{(1)} p_{K^{-}\delta}^{(2)} + \left(p_{K^-}^{(1)} \leftrightarrow p_{K^-}^{(2)}\right), \nonumber\\
    \mathcal{M}_{P}^{\mu} & = & \frac{i}{4} (m_{u} + m_{s}) \frac{Z_{K}}{g_{K}} BW_{K^-} q^{\mu} \left\{g_{\rho}^2 BW_{\rho^0} + g_{\omega}^2 BW_{\omega} + 2g_{\phi}^2 BW_{\phi} \right\}p_{K^-}^{(2)\nu}\left(p_{K^+} - p_{K^-}^{(1)}\right)_{\nu} \nonumber\\
    &&  + \left(p_{K^-}^{(1)} \leftrightarrow p_{K^-}^{(2)}\right),
\end{eqnarray}
where $p_{K^-}^{(1)}$, $p_{K^-}^{(2)}$ and $p_{K^+}$ are the momenta of the final mesons.

The box diagrams have not been taken into account explicitly, because according to our calculations they give a negligible contribution.

The Breit-Wigner propagators describe the intermediate mesons:
\begin{eqnarray}
    BW_{H} = \frac{1}{M_{H}^{2} - p^{2} - i\sqrt{p^{2}}\Gamma_{H}},
\end{eqnarray}
where $H$ designates a meson; $M_{H}$, $\Gamma_{H}$ and $p$ are its mass, width and momentum appropriately.

The variables $h_{A}$ and $h_{V}$ appear in the axial vector and vector channels:
\begin{eqnarray}
    h_{A} & = & M_{K_{1}}^{2} - i\sqrt{q^{2}}\Gamma_{K_{1}} - \frac{3}{2}(m_{s} + m_{u})^{2}, \nonumber\\
    h_{V} & = & M_{K^{*}}^{2} - i\sqrt{q^{2}}\Gamma_{K^{*}} - \frac{3}{2}(m_{s} - m_{u})^{2}.
\end{eqnarray}

Due to the presence of an anomalous type vertex, the following combination of convergent integrals appears in the vector channel:
\begin{eqnarray}
    I_c =I_{21} + m_{u}(m_{s} - m_{u})I_{31},
\end{eqnarray}
where $I_{21}$ and $I_{31}$ take the form (\ref{integral}).

The contributions of the contact diagrams are taken into account in $\mathcal{M}_{A}^{\mu}$ and $\mathcal{M}_{V}^{\mu}$.

The partial width of this decay calculated in the framework of the NJL model takes the value
\begin{eqnarray}
    Br(\tau \to K^- K^+ K^- \nu_\tau) = (2.0 \pm 0.3) \times 10^{-5}.
\end{eqnarray}
The theoretical uncertainty of the used version of the NJL model can be estimated at the level of 15\%.

The experimental value is specified in PDG
\begin{eqnarray}
    Br(\tau \to K^- K^+ K^- \nu_\tau)_{exp} = (2.2 \pm 0.8) \times 10^{-5}.
\end{eqnarray}

\section{Decay $\tau \to K^- K^0 \bar{K}^0 \nu_\tau$}
Due to the isospin symmetry, the width of the process $\tau \to K^- K^0 \bar{K}^0 \nu_\tau$ is expected to be approximately equal to the width of the process $\tau \to K^- K^+ K^- \nu_\tau$. This is confirmed by our calculations. The structure of the amplitude of the process $\tau \to K^- K^0 \bar{K}^0 \nu_\tau$ in the NJL model is similar to the structure of the amplitude of the process $\tau \to K^- K^+ K^- \nu_\tau$ (\ref{amplitude}). The only difference is in the following points: the process $\tau \to K^- K^0 \bar{K}^0 \nu_\tau$ does not contain identical particles, has an additional channel with the charged $\rho$ meson as an intermediate resonance and also contains other relative signs between different contributions. As a result, the axial vector, vector and pseudoscalar channels for this process take the form
\begin{eqnarray}
    \mathcal{M}_{A}^{\mu} & = & \frac{i}{8} (m_{s} + m_{u}) \frac{Z_{K}}{g_{K}} \left[BW_{K_1(1270)} \sin{\alpha} + BW_{K_1(1400)} \cos{\alpha}\right]\left[g^{\mu\nu}h_{A} - q^{\mu}q^{\nu}\right] \nonumber\\
    && \times\left\{\left[g_{\rho}^2 BW_{\rho^0} - g_{\omega}^2 BW_{\omega} - 2g_{\phi}^2BW_{\phi}\right]\left(p_{\bar{K}^0} - p_{K^0}\right)_{\nu} - 2g_{\rho}^2 BW_{\rho^-}\left(p_{K^-} - p_{K^0}\right)_{\nu}\right\}, \nonumber\\
    \mathcal{M}_{V}^{\mu} & = & 2 m_{u} g_{K} I_{c} h_{V} BW_{K^{*-}} \left\{g_{\rho}^{2} BW_{\rho^0} - g_{\omega}^{2} BW_{\omega} + 2 g_{\phi}^{2} BW_{\phi} - 2g_{\rho}^2 BW_{\rho^-}\right\} \nonumber\\
    && \times e^{\mu \nu \lambda \delta} p_{K^-\nu} p_{\bar{K}^{0}\lambda} p_{K^{0}\delta}, \nonumber\\
    \mathcal{M}_{P}^{\mu} & = & \frac{i}{4} (m_{u} + m_{s}) \frac{Z_{K}}{g_{K}} BW_{K^-} q^{\mu} \left\{\left[g_{\rho}^2 BW_{\rho^0} - g_{\omega}^2 BW_{\omega} - 2g_{\phi}^2 BW_{\phi}\right] p_{K^-}^{\nu}\left(p_{\bar{K}^0} - p_{K^0}\right)_{\nu}\right. \nonumber\\
    && \left.- 2g_{\rho}^2 BW_{\rho^-}p_{\bar{K}^0}^{\nu}\left(p_{K^-} - p_{K^0}\right)_{\nu}\right\},
\end{eqnarray}

Its partial width calculated by using the above amplitude takes the following value:
\begin{eqnarray}
    Br(\tau \to K^- K^0 \bar{K}^0 \nu_\tau) = (1.9 \pm 0.3) \times 10^{-5}.
\end{eqnarray}
This result coincides with the result for the process $\tau \to K^- K^+ K^- \nu_\tau$ in the framework of the model uncertainty.

\section{Conclusion}
In the present work, the processes $\tau \to K^- K^+ K^- \nu_\tau$ and $\tau \to K^- K^0 \bar{K}^0 \nu_\tau$ have been considered. Our calculations provided in the framework of the NJL model demonstrate that the main contribution comes from the axial vector channel with the intermediate mesons $K_1(1270)$ and $K_1(1400)$. For example, in the process $\tau \to K^- K^+ K^- \nu_\tau$, the separate contribution of the axial vector channel with the appropriate contact channel is $Br(\tau \to K^- K^+ K^- \nu_\tau)_{K_1} = 1.9 \times 10^{-5}$. The pseudoscalar and vector channels give an order of magnitude smaller contribution in comparison with the axial vector one. The amplitude of the vector channel is orthogonal to the others and does not interfere with them. Besides, it is found that the interference between the pseudoscalar and axial vector channel is negative.

For the decay into charged kaons, a value closer to the result of the experiment \cite{BaBar:2007chl} has been obtained. It is also in good agreement with the middle value specified in PDG \cite{ParticleDataGroup:2022pth}. 

For the decay $\tau \to K^- K^0 \bar{K}^0 \nu_\tau$, a value close to the result for the process $\tau \to K^- K^+ K^- \nu_\tau$ has been obtained. It confirms the suggestion of the approximate isotopic symmetry of this process. Due to the absence of the experimental data for the specified process in PDG, the obtained result can be considered as a prediction.

The theoretical description of the decay $\tau \to K^- K^+ K^- \nu_\tau$ has also been fulfilled in the work \cite{Decker:1992kj}. The vector dominance model with the intermediate resonances is applied there. As a result, the estimation of the partial decay width $5.4 \times 10^{-7}$ is obtained which is significantly lower than the present experimental data. This can be explained by the fact that the important intermediate resonances have not been taken into account in that work. Particularly, the channel containing the $\phi$ meson as an intermediate state has not been considered. However, according to our calculations, it gives the main contribution. Indeed, the channels containing only the $\phi$ meson as a second intermediate state calculated in the framework of the version of the NJL model used in the present work give the value $1.96 \times 10^{-5}$. This channel defines the partial width of the decay $\tau \to K^- K^+ K^- \nu_\tau$ almost completely.

\subsection*{Acknowledgements}
The authors thank prof. A. B. Arbuzov for useful discussions.


\end{document}